\documentstyle[psfig]{mn}

\newif\ifAMStwofonts

\ifoldfss
  \ifCUPmtlplainloaded \else
    \NewTextAlphabet{textbfit} {cmbxti10} {}
    \NewTextAlphabet{textbfss} {cmssbx10} {}
    \NewMathAlphabet{mathbfit} {cmbxti10} {} 
    \NewMathAlphabet{mathbfss} {cmssbx10} {} 
  \fi
  \ifAMStwofonts
    \ifCUPmtlplainloaded \else
      \NewSymbolFont{upmath} {eurm10}
      \NewSymbolFont{AMSa} {msam10}
      \NewMathSymbol{\upi}     {0}{upmath}{19}
      \NewMathSymbol{\umu}     {0}{upmath}{16}
      \NewMathSymbol{\upartial}{0}{upmath}{40}
      \NewMathSymbol{\leqslant}{3}{AMSa}{36}
      \NewMathSymbol{\geqslant}{3}{AMSa}{3E}

       \let\le=\leqslant
       
    \fi
  \fi
\fi 

\ifnfssone
  \newmathalphabet{\mathit}
  \addtoversion{normal}{\mathit}{cmr}{m}{it}
  \addtoversion{bold}{\mathit}{cmr}{bx}{it}
  \newmathalphabet{\mathbfit} 
  \addtoversion{normal}{\mathbfit}{cmr}{bx}{it}
  \addtoversion{bold}{\mathbfit}{cmr}{bx}{it}
  \newmathalphabet{\mathbfss} 
  \addtoversion{normal}{\mathbfss}{cmss}{bx}{n}
  \addtoversion{bold}{\mathbfss}{cmss}{bx}{n}
  \ifAMStwofonts
    \ifCUPmtlplainloaded \else
      \UseAMStwoboldmath
      \makeatletter
      \new@mathgroup\upmath@group
      \define@mathgroup\mv@normal\upmath@group{eur}{m}{n}
      \define@mathgroup\mv@bold\upmath@group{eur}{b}{n}
      \edef\UPM{\hexnumber\upmath@group}
      \new@mathgroup\amsa@group
      \define@mathgroup\mv@normal\amsa@group{msa}{m}{n}
      \define@mathgroup\mv@bold\amsa@group{msa}{m}{n}
      \edef\AMSa{\hexnumber\amsa@group}
      \makeatother
      \mathchardef\upi="0\UPM19
      \mathchardef\umu="0\UPM16
      \mathchardef\upartial="0\UPM40
      \mathchardef\leqslant="3\AMSa36
      \mathchardef\geqslant="3\AMSa3E

       \let\le=\leqslant

    \fi
  \fi
\fi 

\ifnfsstwo
  \DeclareMathAlphabet{\mathbfit}{OT1}{cmr}{bx}{it}
  \SetMathAlphabet\mathbfit{bold}{OT1}{cmr}{bx}{it}
  \DeclareMathAlphabet{\mathbfss}{OT1}{cmss}{bx}{n}
  \SetMathAlphabet\mathbfss{bold}{OT1}{cmss}{bx}{n}
  \ifAMStwofonts
    \ifCUPmtlplainloaded \else
      \DeclareSymbolFont{UPM}{U}{eur}{m}{n}
      \SetSymbolFont{UPM}{bold}{U}{eur}{b}{n}
      \DeclareSymbolFont{AMSa}{U}{msa}{m}{n}
      \DeclareMathSymbol{\upi}{0}{UPM}{"19}
      \DeclareMathSymbol{\umu}{0}{UPM}{"16}
      \DeclareMathSymbol{\upartial}{0}{UPM}{"40}
      \DeclareMathSymbol{\leqslant}{3}{AMSa}{"36}
      \DeclareMathSymbol{\geqslant}{3}{AMSa}{"3E}

       \let\le=\leqslant

    \fi
  \fi
\fi 

\ifCUPmtlplainloaded \else
  \ifAMStwofonts \else 
    \def\upi{\pi}
    \def\umu{\mu}
    \def\upartial{\partial}
  \fi
\fi

\title[Origin of NGC 1404 globular clusters]{Dynamical evolution of globular cluster systems in clusters of galaxies: 
I. The  case of NGC 1404 in the Fornax cluster}
\author[K. Bekki,  Duncan  A. Forbes, Michael A. Beasley,  W.  J. Couch]
       {K. Bekki,${}^1$   Duncan A. Forbes${}^2$, Michael
 A. Beasley${}^2$,  and   W. J. Couch${}^1$\\
        ${}^1$School of Physics, University of New South Wales, Sydney 2052, NSW, Australia \\
        ${}^2$Centre for Astrophysics \& Supercomputing, Swinburne University
of Technology, Hawthorn, VIC,  3122
Australia}
\date{Accepted 
      Received
      in original form 2001}

\pagerange{\pageref{firstpage}--\pageref{lastpage}}
\pubyear{1994}

\begin{document}

\maketitle

\label{firstpage}

\begin{abstract}

We investigate, via numerical simulations, the tidal stripping
and accretion of globular clusters (GCs). 
In particular, we focus on creating models that simulate the
situation for the GC systems of NGC 1404 and NGC 1399 in the
Fornax cluster, which have  poor (specific frequency 
$S_{\rm N}$ $\sim$ 2) and rich ($S_{\rm N}$ $\sim$ 10) GC systems
respectively. 
We initially assign NGC 1404 in our simulation a
typical $S_{\rm N}$ ($\sim$ 5) for cluster ellipticals, and find that
its GC system can only be reduced through stripping  to the presently observed value,
if its orbit is
highly eccentric (with orbital eccentricity of $>$ 0.5) and if
the initial scale length of the GCs system is about twice as large
as the effective radius of NGC 1404 itself. 
These stripped GCs can be said to have formed a `tidal stream'
of intracluster globular clusters (ICGCs) orbiting the
centre of Fornax cluster (many of which would be assigned to NGC
1399 in an imaging study). The physical
properties of these GCs (e.g., number, radial distribution, and 
kinematics) depend on the orbit and 
initial distribution of GCs in NGC 1404.  Our simulations also
predict a trend for $S_{\rm N}$ to rise with increasing
clustercentric distance - a trend for which there is some
observational support in the Fornax cluster. 
We 
demonstrate that since the kinematical properties of ICGCs formed
by tidal stripping in the cluster tidal field depend strongly on
the orbits of their previous host galaxies, observations 
of ICGC kinematics provides a new method for probing 
galaxy dynamics in a cluster.

\end{abstract}


\section{Introduction}

Studies of Globular Cluster (GC) systems have provided  increasingly
useful constraints on the formation and evolution of their host
galaxies (for recent reviews see Harris 2001; Forbes 2002). 
In the case of cD and central cluster galaxies a
variety of processes may have contributed to their rich GC
systems. 
This variety has meant that 
the origin of the rich GC
systems in these galaxies is still poorly understood. 
Ideas to explain the rich GC systems in such galaxies 
have included cooling
flows (Fabian, Nulsen \& Canizares 1984), biased GC formation  (West et
al. 1995), mergers (e.g. Ashman \& Zepf 1992; Bekki et al. 2002), 
tidal stripping and accretion (e.g. Muzzio et
al. 1984; Forbes et al. 1997; C\^ote  et
al. 1998), {\it in situ} formation (e.g. Harris 1991;
Forbes, Brodie \& Grillmair 1997; Harris, Harris \&
McLaughlin 1998). Alternatively 
the galaxies themselves may be `underluminous'
for their GC richness (Blakeslee, Tonry \& Metzger 1997; 
McLaughlin 1999; Beasley et al. 2002). 
It is perhaps the latter three ideas (i.e. accretion, {\it in
situ} formation and missing light) that are considered the
current `best bets' to explain the excess of GCs.

In this paper we further explore the scenario of a central
cluster galaxy accreting GCs via
tidal stripping (as opposed to the accretion of dwarf galaxies
and their GCs as advocated by C\^ote et al. 1998).  
In recent years, a number of new observational facts have lent support
to the stripping and accretion scenario. 
These include: the correlation of 
relative richness of the GC system (called specific frequency S$_N$)
with velocity dispersion (e.g. Blakeslee et al. 1997) - more stripping
should occur in the higher velocity dispersion 
clusters; that high S$_N$
galaxies have relatively more metal-poor GCs (Forbes et
al. 1997) - as metal-poor GCs have a shallower density profile
they are more likely to be stripped from a donor galaxy; 
that S$_N$ values may correlate
with projected clustercentric distance (Forbes et al. 1997) -
galaxies near the cluster core should experience more stripping; 
that the high S$_N$ values in the literature are
generally being revised downwards (see e.g. Forte et al. 2002) 
- this makes the high S$_N$ systems less
extreme and hence less accretion is required.

Although there have been developments on the observational front,
there has been little modelling work carried out since the initial
simulations  by J. C. Muzzio and colleagues in the
1980s (Forte et al. 1982; Muzzio et al. 1984; Muzzio 1986a,b;
Muzzio et al. 1987; Muzzio 1987). They found that the GCs accreted
by the most cluster galaxies do not come from dwarf galaxies
but rather galaxies of similar mass. 
All cluster galaxies gain and lose GCs due to tidal stripping,
and for many galaxies the net outcome was a loss of GCs to the
ICM. When the galaxies represented a small fraction of the total
cluster mass, both the GC gains and losses were smaller.

In the twenty years since the initial models of Muzzio, computing
power has advanced significantly. This combined with further
observational evidence means that the time is ripe to revisit this
issue of GC accretion 
with a N-body simulation. It is however still a formidable task
to perform a fully self-consistent simulation encompassing 
the dynamical evolution of galaxies with vastly different masses
(from small dwarfs to central massive cDs)  and that of GC systems.
Here we focus exclusively on   perhaps the best
case for tidal stripping and accretion, i.e. the GC systems of
NGC~1399 and NGC~1404 in the Fornax cluster.

Several lines of evidence suggest that NGC~1399 is currently in a
non-equilibrium state, and possibly undergoing a dynamical interaction
with its nearby partner NGC~1404.
ROSAT HRI and Chandra data analysed by Paolillo et al. (2001) indicates
that the hot gas centroid of NGC~1399 is offset by $\sim$ 5 kpc to the
SW of its optical counterpart. These authors also find that NGC~1399 has
an extended and asymmetric gas distribution on cluster scales (r
$>$ 90kpc),
which exhibits prominent structures ('holes' and filamentary features)
expected from N-body numerical simulations (e.g. Barnes 2000).

Independent evidence comes from discrete dynamical tracers such as
GCs (Grillmair etal. 1994; Minniti etal. 1998; Kissler-Patig etal. 1999)
and planetary nebulae (PNe, Arnaboldi etal. 1994; Napolitano et al. 2002).
In particular, Napolitani et al. (2002) find evidence for a disturbed velocity
structure after re-analysing the PNe data of Arnaboldi etal.
Their non-equilibrium dynamical analysis shows a peak rotation of
250 km s$^{-1}$ for the PNe at $\sim$ 12.6 kpc from the optical centre of
NGC~1399, coincident with a
velocity dispersion minimum at this point, which subsequently rises at
larger radii. A viable analytical explanation for this kinematical
behavior is that NGC~1404 has undergone a "flyby" of NGC~1399, whose
stellar halo is currently attempting to return to equilibrium (Napolitani
et al. 2002).

Thus the purpose of this paper is to investigate numerically
whether tidal stripping of the NGC~1404 GC system by the global
gravitational field of the Fornax cluster (and thus NGC 1399) can
explain physical properties of its GC system. In particular, we 
investigate (1) whether the observed relatively low S$_N$ value
of 2 in NGC 1404 can be explained by the tidal stripping of GCs
and (2) how the density profile of the GC system can change after
GC stripping.  
We also try to identify observational signatures
of the GC stripping and accretion process. 
White (1987) suggested that intracluster GCs can be formed
owing to tidal stripping of GCs from cluster member galaxies
and Bassino et al. (2002) 
have recently found GC candidates in the Fornax cluster.
In light of these results, we discuss the origin of intracluster GCs in
the Fornax cluster in the context of GC stripping from NGC 1404.
This paper is the first
in a series of papers in which we seek a 
self-consistent understanding of dynamical
evolution of GC systems in cluster environments.
Here we focus on the limited case of the interaction between NGC
1399 and NGC 1404, and its effect on their GC systems.  
The distance of the Fornax cluster is 
assumed to be 18.6 Mpc throughout this paper. 

\begin{figure*}
\psfig{file=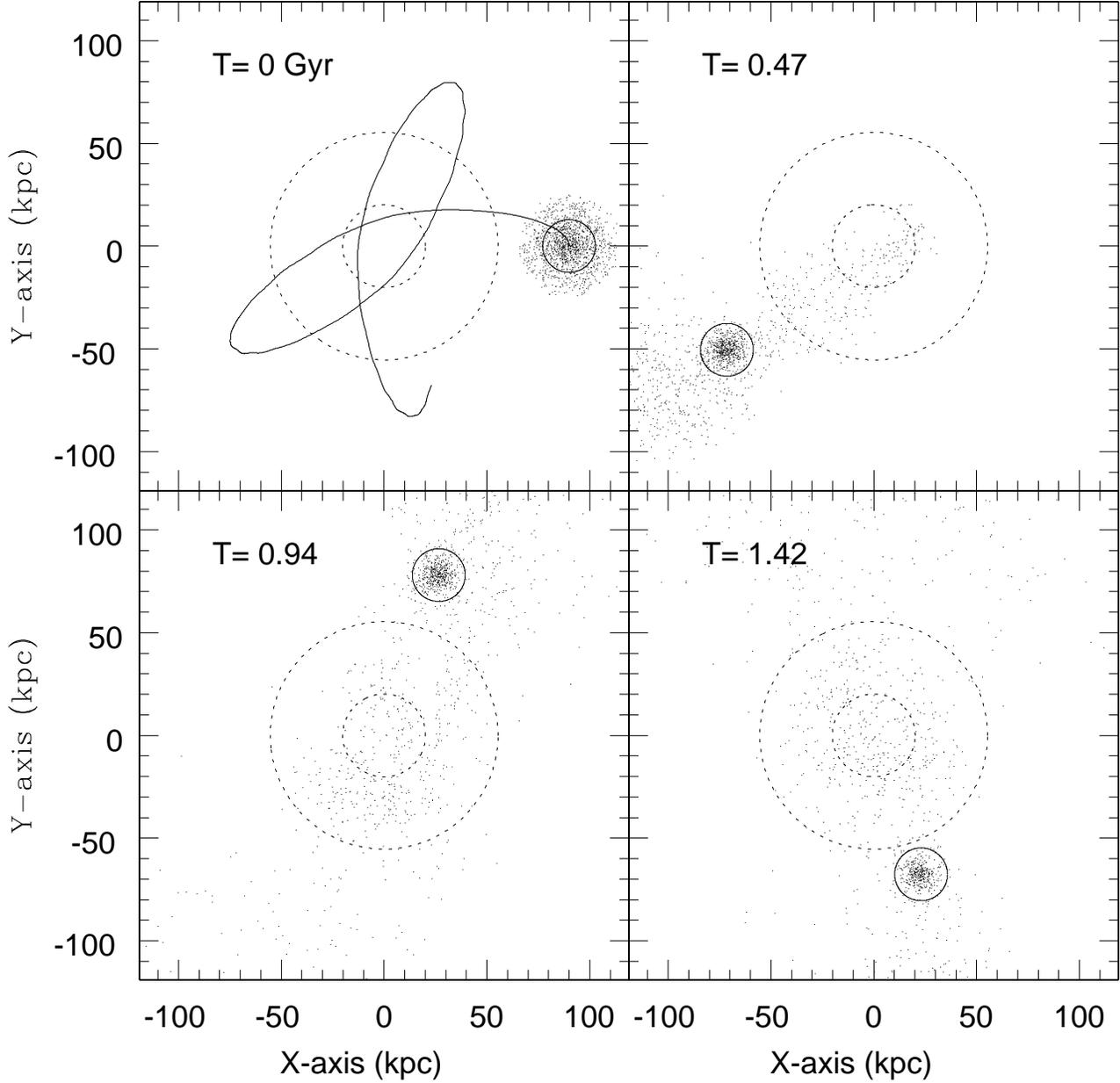}
\caption{ 
Morphological evolution of the globular cluster system of NGC
1404 orbiting within the Fornax cluster.  here we show the fiducial
model (Model 3) with $e_{\rm p}$ = 0.76 and $a_{\rm gc}$ = 2.0
projected onto the $x$-$y$ plane.  The time $T$ (in Gyr)
represents the time that has elapsed since the simulation
started. The larger and smaller dotted circles represent the
cluster scale radius $r_{\rm s}$ of the adopted NFW mass profile
and 5 $R_{\rm e}$ (where $R_{\rm e}$ is effective radius) of the
central NGC 1399, respectively. Solid lines represent the orbit
of NGC 1404 (for 0 $\le$ $T$ $\le$ 1.42 Gyr) and 5 $R_{\rm e}$ of
NGC 1404.  }
\label{Figure. 1}
\end{figure*}

\section{The model}

We consider a collisionless stellar system of an elliptical
galaxy with a mass and size similar to those of NGC 1404,
orbiting the centre of Fornax cluster of galaxies (i.e., NGC
1399). Since the gravitational field around NGC 1399 at distances
greater than 40 kpc is observed to be dominated by the global
cluster rather than by luminous components of NGC 1399 itself
(e.g., Kissler-Patig et al. 1999; Napolitano et al. 2002), we
consider that the gravitational field of the dark matter halo of
the cluster has the strongest influence on the dynamical evolution
of the globular cluster system in NGC 1404.  Accordingly, we
model only the cluster tidal field and do not include any tidal
effects of other cluster member galaxies in the present
simulations. Although these simulations are idealized 
in some aspects, we believe that our model still contains the 
essential ingredients for the dynamical evolution of
NGC 1404's globular cluster system in Fornax cluster.

To give our model a realistic radial density profile for
the  dark matter halo, we base it on both the X-ray observational
results of Jones et al. (1997) and the predictions from the standard cold
dark matter cosmogony (Navarro, Frenk, \& White 1996, hereafter NFW). 
The NFW profile is described as,
\begin{equation}
{\rho}(r)=\frac{\rho_{0}}{(r/r_{\rm s})(1+r/r_{\rm s})^2},
\end{equation}
where  $\rho_{0}$ and $r_{\rm s}$ are the central density and the scale
length of a dark halo, respectively.
$\rho_{0}$ is chosen such that the total mass of
Fornax cluster  within 125 kpc  is the same as 
the observed one  ($8.1 \times 10^{12} \rm M_{\odot}$)
estimated from X-ray observations  (Jones et al. 1997).
We adopt 55 kpc for $r_{\rm s}$ which is reasonable
value for a CDM halo with the mass similar to that of 
Fornax cluster.

NGC 1404 is modeled as a fully self-gravitating system
and assumed to consist of dark matter halo, stellar component, and
GCs. The density profile of the dark  matter halo 
of the elliptical is also represented by NFW profile with the scale 
length the same as the effective radius of the stellar component
of NGC 1404 (i.e. 2.5 kpc). The mass ratio of the dark matter halo
to the stellar component is set to be 5. The effective radius and
the total mass of the stellar components is 2.5 kpc and 
4.4 $\times$ $10^{10}$ $M_{\odot}$, respectively,
which are consistent with observational properties of NGC 1404. 
The projected density profile of the stellar component
is represented by $R^{1/4}$ law and the density profile is cut off
at $R$ = $5R_{\rm e}$.  
We estimate both  the velocities of the dark matter halo particles and 
those of stellar ones from the gravitational potential
at the positions where they are located.
In detail, 
we first calculate the one-dimensional isotropic dispersion according to
the (local) virial theorem:
\begin{equation}
{\sigma}^{2}(r)=-\frac{U(r)}{3},
\end{equation}
where $U(r)$ is the gravitational potential at the position $r$.
Then we allocate a velocity to  each collisionless particle (dark matter
halo and stellar particles) so that the distribution of velocities
of these particles can have a Gaussian form with a dispersion equal
to ${\sigma}^{2}(r)$. 

It should be noted here that we use the above $U(r)-{\sigma}^{2}(r)$ relation
rather than the  Jeans equation 
for a spherically symmetric system (Binney \& Tremaine 1987);
\begin{equation}
\frac{d(\rho (r){\sigma}^{2}(r))}{dr}=-\rho (r) \frac{d \Phi (r)}{dr}.
\end{equation}
This is firstly because the self-gravitating systems modeled in the present
study are composed of three different stellar components (dark matter halo,
$R^{1/4}$-law elliptical stellar component, and GCs) with non-analytical
radial density distributions and secondly because we introduce the cut off 
radius for each component.  
The derived ${\sigma}^{2}(r)$ at each location of a particle can be 
consistent with those derived in the Jeans equation.

In the present study, we introduce the cut off radius ($R_{\rm c}$)
beyond which no dark matter (stellar) particles are initially allocated
($R_{\rm c}$ = $5R_{\rm e}$). Because of this somewhat artificial 
boundary condition, the very outer part of the NGC 1404 system
is initially not in a dynamical equilibrium in a strict sense. 
Therefore we need to make the dark matter component dynamically relaxed after
giving ${\sigma}(r)$ to each dark matter particle
and then   use the relaxed system as an initial model for
each simulation.
We ran an isolated elliptical model for five dynamical time
scales until the dark matter component reaches  the new dynamical
state (during this dynamical relaxation, the outer part of the system 
expands only slightly). For all simulations,
this dynamically relaxed  model is used as an initial NGC 1404 system.

The  globular cluster distribution is assumed to follow
$\rho (R)$ $\propto$ ${(R_{\rm c}+R)}^{\alpha}$, where $R$,
$R_{\rm c}$, and $\alpha$ are the radius from the centre of NGC
1404, the core radius of the GC distribution, and the slope of
the GC profile.  There are only a few papers which have attempted
to derive the slope of the GC distribution both for metal-poor
GCs and for metal-rich GCs (e.g. Geisler et al. 1996), though a
number of observational studies have estimated the profiles for
the {\it whole} GC populations and their correlations with other
global properties of GC's host galaxies (e.g., Forbes et
al. 1997). In the present paper, we do not distinguish blue
metal-poor GCs and red metal-rich ones. Therefore we use the
observed mean value of $\alpha$ (= $-1.9$) derived for the whole
GC system  for the projected density profile of the GC system
of NGC 1404.  However we note that the metal-poor GCs have a
shallower distribution (Forbes et al. 1998).

The ratio of $R_{\rm c}$ to $R_{\rm e}$ is
considered to be an important free parameter and represented by
$a_{\rm gc}$. The total number of GCs is set to be 1350, which
corresponds to initial $S_{\rm N}$ of 5.0 (i.e. a typical cluster
elliptical value). Each GC has a mass of
$10^6$ $M_{\odot}$ and its velocity is given in the same way as
each stellar (dark matter) particle is given its velocity.
The present numerical results do not depend  strongly
on the mass of each GC particle ($M_{\rm gc}$),
and the dependences on $M_{\rm gc}$ are briefly summarized
in the Appendix A for $10^5$ $M_{\odot}$ $\le$ $M_{\rm gc}$  $\le$ $10^6$ $M_{\odot}$.

The orbit of NGC 1404  is assumed to be influenced
only by the gravitational potential resulting from the dark halo
component of the Fornax cluster. Since the adopted cluster potential
is spherical symmetry (not triaxial), 
the orbit of NGC 1404 forms a rosette within a plane. 
This orbital plane is set to be  $x$-$y$ in all models. 
The centre of the cluster  is 
always set to be ($x$,$y$,$z$) = (0,0,0) whereas the initial position
of NGC 1404 is set to be ($x$,$y$,$z$) = ($R_{\rm ini}$, 0, 0).
The projected distance between NGC 1399 (or the centre of Fornax cluster)
and NGC 1404 (hereafter $R_{\rm obs}$)
and the relative radial velocity of the two galaxies are  observed to
be $\sim$ 45 kpc and 525 km s$^{-1}$, respectively (Forbes et al. 1997).
Since these observational values can not give any strong constraints on the orbit
of NGC 1404, 
we consider any reasonable  values of $R_{\rm ini}$ larger than $R_{\rm obs}$.
We show the results of the models with  $R_{\rm ini}$ = 2 $R_{\rm obs}$ (90 kpc)
in which the projected distance between the two galaxies can be the same as the observed
one (depending on the line of sight) during orbital evolution of these models.

The initial velocity of NGC 1404 ($v_{\rm x}$,$v_{\rm y}$,$v_{\rm
z}$) is set to be (0, $f_{\rm v} V_{\rm c}$, 0), where $f_{\rm
v}$ and $V_{\rm c}$ are the parameters controlling the orbital
eccentricity (i.e, the larger $f_{\rm v}$ is, the more circular
the orbit becomes) and the circular velocity of the cluster at
$R$ = $R_{\rm ini}$, respectively.  We investigate three
representative values of $f_{\rm v}$ = 0.25, 0.5, and 1, which
corresponds to NGC 1404's orbital eccentricity (hereafter $e_{\rm
p}$) of 0.76, 0.5, and 0.0 (circular), respectively.

Two parameters $a_{\rm gc}$ and $e_{\rm p}$ (or $f_{\rm v}$) are
considered to be the most important parameters for the dynamical
evolution of NGC 1404's GCs 
in the present study: These determine the effectiveness of tidal
stripping of GCs in a cluster environment.  In total, we show the
results of the following 6 models: Model 1 with $e_{\rm p}$ = 0.0
and $a_{\rm gc}$ = 2.0, 2 with $e_{\rm p}$ = 0.50 and $a_{\rm
gc}$ = 2.0, 3 with $e_{\rm p}$ = 0.76 and $a_{\rm gc}$ = 2.0, 4
with $e_{\rm p}$ = 0.00 and $a_{\rm gc}$ = 1.0, 5 with $e_{\rm
p}$ = 0.50 and $a_{\rm gc}$ = 1.0, and 6 with $e_{\rm p}$ = 0.76
and $a_{\rm gc}$ = 1.0.  We describe the results of model 3
(hereafter referred to as a fiducial model) in more detail,
because this model shows  behaviours typical of dynamical evolution
of the GC system in our models. All the simulations have been
carried out on the GRAPE board (Sugimoto et al. 1990) with the
total particle number of 11350 (10000 for dark matter and stellar
particles and 1350 for GC particles) and the gravitational softening
length of each component equal to the mean particle separation at
the half mass radius of each component (i.e., 0.71, 0.55, and
2.95 kpc for dark matter, stars, and GCs, respectively).

\section{Results}

\subsection{Fiducial model}

Fig. 1 summarises the dynamical evolution of GCs in NGC 1404
orbiting the central dominant galaxy (NGC 1399) of Fornax
cluster. As the galaxy passes through the cluster core for the
first time, GCs in NGC 1404 are efficiently
stripped owing to the strong cluster tidal field. About 46 \% of the GCs
are located outside 10 $R_{e}$, where $R_{e}$ is the effective
radius of NGC 1404, and we regard them as being tidally stripped
after this first pericenter passage ($T$ = 0.47 Gyr), whereas
only $\sim$ 6 \% of field stars  are
stripped owing to its their more compact configuration. The stripped
GCs are distributed along the orbit of NGC 1404 and some fraction
of them can be clearly seen within the NGC 1399's optical radius
(defined as 5 $R_{\rm e}$ of NGC 1399 in this paper). Since the
orbit of NGC 1404 in the adopted spherical gravitational
potential forms a rosette, the stripped GCs can be finally
distributed widely throughout the cluster scale radius $r_{\rm
s}$ ($\sim$ 55 kpc). The overall (projected) distribution of the
stripped GCs within the central 100 kpc of the cluster can be
regarded as rather elongated, which reflects the fact that the
NGC 1404's orbit is highly eccentric (orbital eccentricity
$e_{p}$ = 0.76). The overall distributions also show appreciable
inhomogeneity, in particular, for the regions with $R$ $>$ 50
kpc.

\begin{figure}
\psfig{file=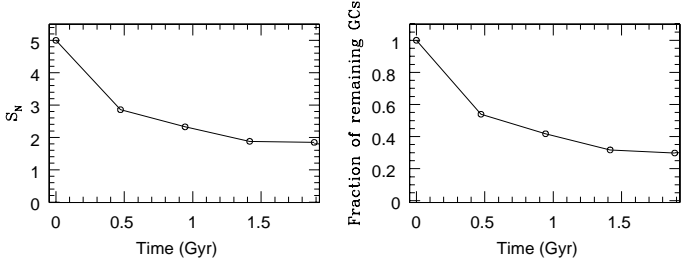,width=9.0cm}
\caption{ 
Time evolution of $S_{\rm N}$ (left) and number of GCs remaining 
within  NGC 1404 (right) in the fiducial model. 
NGC 1404 is assumed to have 1350 GCs initially within 10 $R_{\rm e}$. 
}
\label{Figure. 2}
\end{figure}

During dynamical evolution of NGC 1404 within the cluster, the
GCs are more strongly influenced by the  cluster tidal
field than the main stellar component of NGC 1404, because the GC
system is initially less compact.  Fig. 2 shows how this
difference in the effectiveness of tidal stripping between the
stellar component and GCs can cause the  decrease of
$S_{\rm N}$ in NGC 1404.  After the first pericenter passage, the
$S_{\rm N}$ is reduced from 5 to 2.86 ($T$ = 0.47), and the
reduction rate is nearly proportional to the number of the
stripped GCs.  GCs can be successively stripped everytime NGC
1404 passes through the cluster core, though the total number of
the stripped ones become smaller in the second and the third
pericenter passages compared with those stripped in the first
pericenter passage ($S_{N}$ = 2.32 and 1.88 for $T$ = 0.94 and
1.42, respectively).  These results clearly suggest that the
origin of the observed smaller $S_{N}$ ($\sim$ 2) of NGC 1404 GCs
may be closely associated with tidal stripping of GCs by the cluster
tidal field.  Finally, the dynamical evolution of these stripped
GCs becomes gravitationally influenced by the cluster potential
rather than by NGC 1404 and therefore can be identified as
drifting intracluster GCs.

The NGC 1404's tidal radius ($R_{\rm t}$), within which collisionless components
(dark matter, stellar components, and GCs) do not suffer so severely from
tidal stripping by the cluster,
can be analytically estimated to be 3.7 $R_{\rm e}$ for 
the fiducial model with $e_{\rm p}$  = 0.76 and the total
mass of NGC 1404 equal to 2.64 $\times$ $10^{11}$ $M_{\odot}$. 
The initial total number of GCs within $R_{\rm t}$ is 
288 for the fiducial model, and accordingly 1062 GCs are expected to be tidally  stripped
from NGC 1404.  
The final number of GCs within $R_{\rm t}$ is 253 ($T$ = 1.42 Gyr) in
the simulation, and accordingly 1097 GCs are tidally  stripped  from NGC 1404.
Therefore, the simulated number of GCs stripped from NGC 1404 
differs from the analytical estimation only by
3 \%. The probable reason for the larger number of the stripped GCs in the simulation
is that the NGC 1404 system becomes less strongly self-gravitating
(thus more susceptible to tidal stripping) after stripping
of dark matter and stellar components during its dynamical evolution.
The possibility of the present model's overestimation of the number of stripped GCs
(due to numerical relaxation effects) is discussed in  the Appendix B.

Fig. 3 shows that as NGC 1404 orbits the centre of 
the Fornax
cluster, the radial distribution of GCs in NGC 1404 becomes
steeper and the radial $S_{N}$ distribution becomes flatter. This
is essentially because GCs initially outside the galaxy are more
likely to be tidally stripped.  The $S_{N}$ within 5$R_{\rm e}$
is changed from 2.83 to 1.64 (a factor of 1.73 smaller) at $T$ =
1.42 Gyr whereas the $S_{N}$ within 10$R_{\rm e}$ is changed from
5 to 1.88 (a factor of 2.66 smaller). These results suggest that
tidal stripping of GCs causes the steepening of radial
distributions of GCs. Furthermore, they imply that if a cluster
elliptical galaxy with lower $S_{N}$ has a typical $S_{N}$
estimated for its halo region within a few times $R_{\rm e}$ and
a significantly  smaller $S_{N}$ within 5 $-$ 10 $R_{\rm e}$, the
lower $S_{N}$ of this elliptical can be caused by the strong cluster
tidal field. Thus, we  suggest that the ratio of the inner $S_{N}$
and the outer $S_{N}$ in a cluster elliptical is one observable
test for the tidal stripping of GCs in 
cluster elliptical galaxies.

The derived radial density profile with the power-law slope of
$\sim$ $-2.6$ for 0 $\le$ $R/R_{\rm e}$ $\le$ 10 is steeper than
the observed value of $-1.3$ (Forbes et al. 1997).  This
disagreement between the present simulations and the observation
is due partly to the adopted initially steep slope ($-1.9$),
which is chosen such that the value is similar to the typical
value of the power-law slope of globular cluster systems in
elliptical galaxies. As the simulations demonstrate, the density
profile of the GC system of an elliptical in a cluster is more
likely to become steeper as it orbits the cluster because of the
more efficient tidal stripping in the outer part of the galaxy.
We therefore suggest that if the origin of the observed low
$S_{\rm N}$ of NGC 1404  is due to tidal stripping, the
power-law slope of the GCs radial density profile should have been
initially flatter (i.e. before interaction with Fornax cluster).

\begin{figure}
\psfig{file=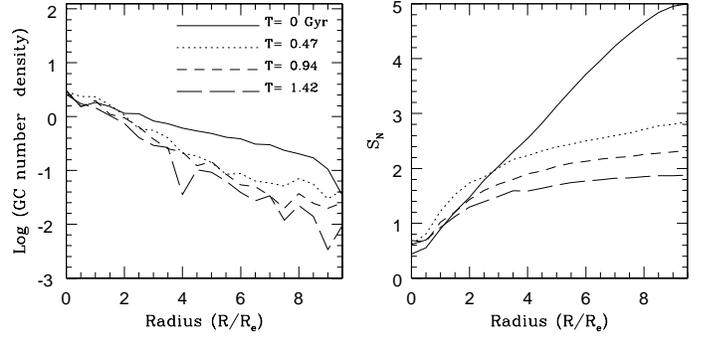,width=9.cm}
\caption{ 
Time evolution of projected surface number density of NGC 1404 GCs
(left) and the radial dependence of $S_{\rm N}$  (right) in the fiducial model.
It should be noted that the decrease  of $S_{\rm N}$ is more significant 
in the outer parts of the galaxy.
}
\label{Figure. 3}
\end{figure}

\begin{figure}
\psfig{file=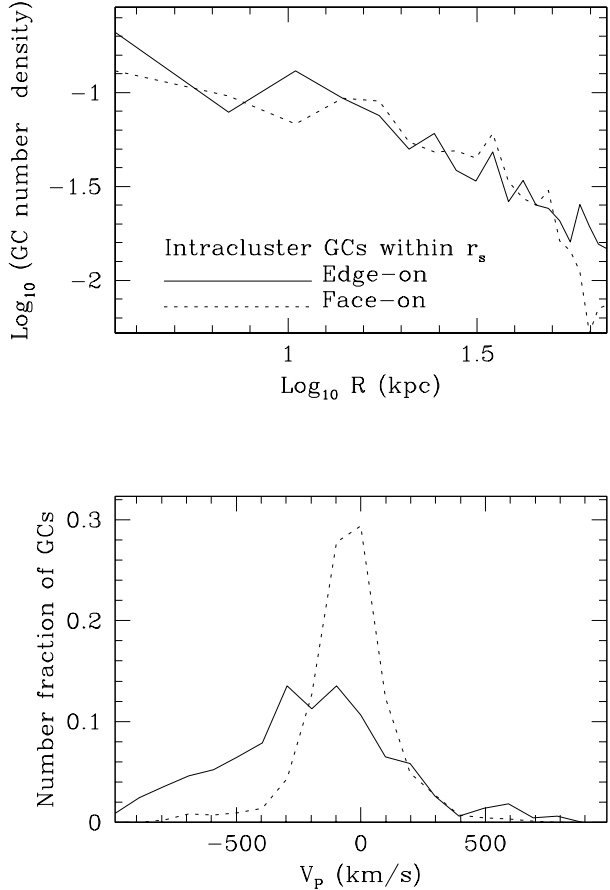,width=8.cm}
\caption{ 
$Upper$: Final (projected) surface number distribution of intracluster globular clusters (ICGCs)
that are within $r_{\rm s}$ of the Fornax cluster model  and
were stripped from NGC 1404 
for the fiducial model (Model 3) at $T$ = 1.9 Gyr. Here the distance ($R$) of
each ICGC is that from the centre of the cluster (not from the centre of NGC 1404).
Both vertical and horizontal scales are given in log scales and solid and dotted
lines represent the result for the edge-on  view (seen from the orbital plane
of NGC 1404) and for the face-on one (from the above of the orbital plane),
respectively. 
$Lower$: Final histogram of projected velocity dispersion of the ICGCs
for the edge-on  view  (solid) and the face-one one (dotted). 
}
\label{Figure. 4}
\end{figure}

\begin{figure}
\psfig{file=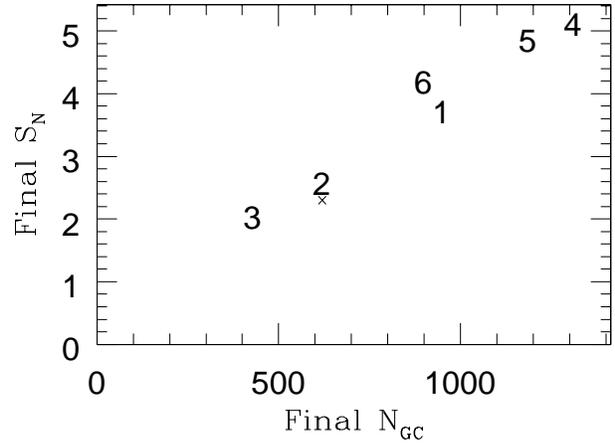,width=8.cm}
\caption{ 
Final distribution of the six  models (Model 1 - 6)
with different orbital configurations and initial GC distributions
in the  $S_{\rm N}$-$N_{\rm GC}$ plane.
Model 1 ($e_{\rm p}$ = 0.00 and $a_{\rm gc}$ = 2.0), 
2 ($e_{\rm p}$ = 0.50  and $a_{\rm gc}$ = 2.0), 
3 ($e_{\rm p}$ = 0.76  and $a_{\rm gc}$ = 2.0), 
4 ($e_{\rm p}$ = 0.00  and $a_{\rm gc}$ = 1.0), 
5 ($e_{\rm p}$ = 0.50  and $a_{\rm gc}$ = 1.0), 
and 6  ($e_{\rm p}$ = 0.76  and $a_{\rm gc}$ = 1.0) are represented by
their numbers.
For comparison, the observed values of $S_{\rm N}$ and $N_{\rm GC}$ of NGC 1404
is plotted by cross.
}
\label{Figure. 5}
\end{figure}

The radial number density profile of the GCS around NGC 1404 can be influenced
by later tidal disruption of GCs by NGC 1404. We can roughly
estimate the effect of tidal disruption of GCs on the 
number density profile by using early numerical results by Aguilar et al. (1988).
They demonstrated that if the Galactic GCs are within $\sim$ 2 kpc from the
Galactic centre, the GCs can be destroyed by the strong tidal field of the bulge.
The radius within which GCs can be destroyed is referred to as $R_{\rm des}$ 
from now on for convenience ($R_{\rm des}$ $\sim$ 2 kpc for the Galactic bulge).
By assuming that $R_{\rm des}$ is simply scaled  to ${M_{\rm gal}}^{1/3}$
for a galaxy,
(where $M_{\rm gal}$ is the total luminous mass of the galaxy),
$R_{\rm des}$ for NGC 1404 can be estimated to be 3.3 kpc (or 1.3 $R_{\rm e}$).
In this estimation, we consider that GCs can be destroyed 
by the more compact luminous component rather than by the diffuse dark matter halo
in the central region of NGC 1404. Therefore, 
we suggest that since   GCs within 1.3 $R_{\rm e}$ can be  all
destroyed by the NGC 1404's tidal field,  
the radial density profile of the NGC 1404's GCS can become
more flattened for $R/R_{\rm e}$ $<$ 1.3 (compared with the simulated distribution).
The outer density distribution of GCs ($R/R_{\rm e}$ $>$ 2) 
could not be influenced by the GC disruption.

Fig. 4 shows projected number density profiles of ``stripped''
GCs {\it with respect to the centre of the cluster} at $T$ =1.42
Gyr.  Here the ``stripped'' GCs are defined as those that have a
projected distance (from NGC 1404) of larger than 10 $R_{\rm e}$
(where $R_{\rm e}$ is the effective radius of NGC 1404).  It
should be stressed here that the adopted radius of 10 $R_{\rm e}$
just corresponds to the initial size of the GC system and thus
some of GCs outside this radius could be still orbiting NGC 1404
(i.e., not all of GCs outside this radius are really stripped
from NGC 1404 gravitational field). Since it is very difficult to
determine, only from {\it projected radius and kinematics} in the
simulations, whether a GC is really stripped from NGC1404, we
consider that this adopted 10 $R_{\rm e}$ is a plausible value
which gives a boundary between GCs trapped in NGC 1404 and those
stripped. We suggest that these stripped GCs (far from their
previous host galaxies) could be identified as ``intracluster
globular clusters'' (defined by West 1990) in future observations
studies. These intracluster globular clusters are referred to as
ICGCs in the present study, however we emphasize that they did
not form in the ICM but now reside there.

The derived number density profiles of ICGCs within the cluster
scale radius $r_{\rm s}$ clearly show negative gradients both in
edge-on and face-on views. The power-law slopes of these can be
estimated to be $\sim$ $-0.85$ within $r_{\rm s}$, which is
similar to the observed slope ($-1.0 \pm 0.2$) of the central
metal-poor GCs in NGC 1399 (e.g., Forbes et al. 1997; Also see
the Fig.2 in Hilker et al. 2000) and rather shallower than that
of the metal-rich ones ($-1.7\pm 0.2$; See Dirsch et al. 2003 for
the latest value).  One possible reason
for the negative gradient in the ICGC distribution is that GCs
are more likely to be stripped in the inner regions of the
cluster. The derived value of the slope implies that only the
blue GCs in the central region of Fornax cluster (equally, GCs of
NGC 1399) were previously NGC 1404's GCs. Blue GCs in elliptical
galaxies show more extended distributions or
shallower density profiles than red GCs (e.g., Geisler et al
1996; Forbes, Brodie \& Huchra 1997).  
Our simulations demonstrate that the outer GCs are more
likely to be stripped during the dynamical evolution than inner
ones.  Thus, although GCs stripped from NGC 1404 consists of only
some fraction of GCs in the central region of Fornax cluster, it
is reasonable to claim that the tidal stripping processes of {\it
outer blue} GCs from NGC 1404 can contribute partly to the
observed power-law slope of blue GCs in the central region of
Fornax cluster.

Fig. 4 furthermore demonstrates that the ICGCs show wider wings
of the (projected) velocity ($V_{\rm p}$) distribution in the
edge-on view whereas they have the sharp distribution in the
face-one one. This result just reflects that the edge-on view is
nearly coincident with the orbital planes for most of drifting
ICGCs. A larger fraction of large projected velocity of ICGCs
with respect to the cluster centre ($>$ 400 km s$^{-1}$) can be
found in the edge-on distribution, in particular, in its negative
velocity side.  The velocity dispersion for ICGCs within $r_{\rm
s}$ (and with the absolute magnitude of projected velocity
$V_{\rm p}$ less than 1000 km s$^{-1}$) can be estimated to be
340 km s$^{-1}$ for the edge-on view and 174 km s$^{-1}$ for face-one
one.  Some of ICGCs with $|V_{\rm p}|$ $>$ $400$ km s$^{-1}$ are
located close to the central NGC 1399 (i.e. within 5 times the
effective radius of NGC 1399) so that these can be identified as
NGC 1399's GCs (not as ICGCs) with an unusually large projected
velocity.

The derived velocity dispersion of 340 km s$^{-1}$ of ICGCs, some
fraction of which are close to the centre of the cluster (i.e.,
NGC 1399), is significantly larger than the observed outer
stellar velocity dispersion ($\sim$ 200 km s$^{-1}$ for the
region
1 -  8 kpc  from the centre) in NGC 1399 (e.g., Bicknell et al. 1989).
Based on the radial velocity measurement of 74 GCs around NGC 1399,
Kissler-Patig et al. (1999) found that the velocity dispersion
for the whole sample is $373 \pm 35$  km s$^{-1}$ and the velocity
dispersion depends on radius in such a way that it {\it increases} with
radius from the NGC 1399 centre between 10 - 30 kpc (e.g., $263
\pm 92$ at 2 $\acute{}$ and $408 \pm 107$ km s$^{-1}$ at 5
$\acute{}$ ).  This outer increase of velocity dispersion has
also been found in the kinematical analysis of planetary nebulae
(e.g., Napolitano, Arnaboldi, and Capaccioli 2002).  The observed
velocity dispersion of GCs (and PNe) is larger than the (outer)
stars of NGC 1399, which is considered to be a possible evidence
that GCs (PNe) are orbiting in a gravitational potential of the
Fornax cluster rather than in NGC 1399 itself (Grillmair et
al. 1994; Kissler-Patig et al. 1999).

Our results on the larger velocity dispersion of ICGCs strongly
suggests that if they originated  from tidal stripping of GCs
initially in NGC 1404 and observationally identified as GCs in
NGC 1399, these GCs can have a velocity dispersion as large as
the observed one for GCs in the very outer part of NGC
1399. These results furthermore imply that if outer GCs in NGC
1399 are composed mostly of ICGCs stripped from other cluster
member galaxies, GCs in NGC 1399 as a whole can show a clear
difference in velocity dispersion between the inner ``intrinsic''
GCs that are closely associated with the NGC 1399 formation
itself and the outer accreted/stripped ones from other cluster
member galaxies. Our numerical simulations have been carried out
{\it only} for the dynamical evolution of NGC 1404 with plausible
orbits. The (projected) radial velocity dispersion profile of
ICGCs depends on from which galaxy these are stripped, because
the projected velocity of each stripped GC depends on the orbit
of its previous host galaxy (i.e., the orbit of each GC follows
the orbit of its host).  We thus suggest that the observed radial
velocity dispersion profile of GCs in NGC 1399 may be ``a fossil
record'' of the histories of GC stripping from other cluster
member galaxies around NGC 1399. However it may be difficult in
practice to disentangle kinematically any stripping signature
from other (formation) mechanisms that give rise to the rich GC
systems of cD galaxies.

\subsection{Parameter Dependences}

Although our general results on the stripping of GCs from NGC
1404 due to the strong tidal field of the Fornax cluster do not depend on
the model parameters, the final $S_{\rm N}$ and $N_{\rm GC}$ do
depend on (1) orbital eccentricity ($e_{\rm p}$) of the host
galaxy NGC 1404 and (2) initial ratio of the scale length of GC
system to the effective radius of NGC 1404 ($a_{gc}$). The final
$S_{\rm N}$ and $N_{\rm GC}$ of GCs around NGC 1400 are estimated
at $T$ = 1.89 Gyr when time evolution of $S_{\rm N}$ and $N_{\rm
GC}$ becomes insignificant for all models (Model 1 - 6). We can
only discuss the final structural and kinematical properties of
ICGCs for the models 2 and 3, in which total number of GCs are
large enough.  (Total number of stripped GCs in Model 1, 4, 5,
and 6 are too small for us to derive velocity dispersion of ICGCs
at a given radius). Therefore, we only briefly describe the
dependences of ICGC properties on model parameters.  In Figs. 5
and 6, we illustrate the derived dependences on the above two
parameters.  We find the following:

(i) Both $S_{\rm N}$ and $N_{\rm GC}$ are smaller for the models with 
larger $e_{\rm p}$, because in the models with larger $e_{\rm p}$
(i.e., with smaller pericenter distance),
galaxies can pass through the inner region of the cluster, where the cluster tidal  
field is sufficiently strong such  that a larger number of GCs can be stripped.
This dependence does not depend on $a_{\rm gc}$.

(ii) Irrespective of  $e_{\rm p}$ and $a_{\rm gc}$, 
the model with smaller $N_{\rm GC}$  shows smaller $S_{\rm N}$. 
This means that the stripping of stars in NGC 1404, which can increase
$S_{\rm N}$ if stripping of GCs does not happen, 
is much less efficient compared with GC stripping,
so that  the decrement of  $S_{\rm N}$ depends strongly on 
that of $N_{\rm GC}$.

(iii) Final $S_{\rm N}$ depends on $a_{\rm gc}$  such that
models with smaller $a_{\rm gc}$ have larger final  $S_{\rm N}$
because of its less efficient tidal stripping of GCs. Comparison
of our results and the observations suggests that in order to explain
both the observed $N_{\rm GC}$ and $S_{\rm N}$,
{\it both $e_{\rm p}$ and  $a_{\rm gc}$ should be large}.
(Either  Model 2 or  3  can best reproduce the observed properties 
of NGC 1404 in the present study).

(iv) Our models predict that $S_{\rm N}$  correlates with the distance 
of GCs' host galaxy from the centre of Fornax cluster (equivalently,
from  NGC 1399) in such a way that a galaxy with the larger distance
shows larger $S_{\rm N}$. 
In Fig 7 we plot the observed SN of Fornax
ellipticals as a function of cluster-centric radius. These data suggest
that galaxies at larger radii have higher SN. 
The location of NGC 1404 GCs in Fig. 6 also
suggests that if the observed lower $S_{\rm N}$ and $N_{\rm GC}$ in NGC 1404 are 
due to tidal stripping by the cluster tidal field, 
$e_{\rm p}$  should be as large as (or larger than) 0.5 and
$a_{\rm gc}$ is $\sim$ 2.

(v) Projected surface number density profiles and velocity dispersion of ICGCs 
(around  NGC 1399) are likely to be shallower and smaller, respectively,
in the model with smaller $e_{\rm p}$, though we could investigate these
only for the models with  $a_{\rm gc}$ = 2.0 because of the very smaller
number of ICGCs in other models. These results indicate that
physical properties of ICGCs in a cluster depends on the  orbital populations
(e.g., the mean orbital eccentricity) of  galaxies in the cluster. 


\begin{figure}
\psfig{file=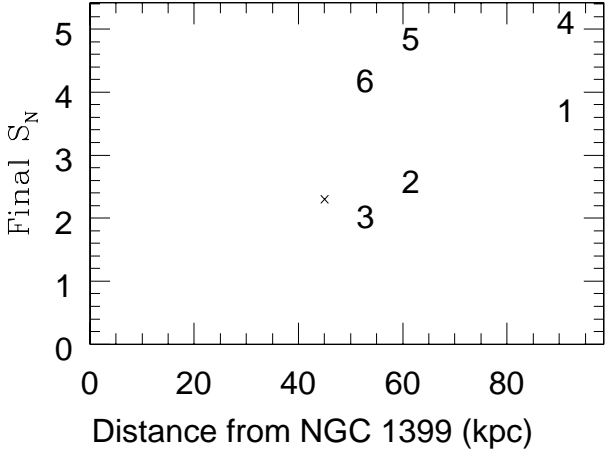,width=8.cm}
\caption{
Variation of $S_{\rm N}$ with projected distance.
Final distribution of the six  models (Model 1 - 6)
with different orbital configurations and initial GC distributions
on a plane defined by $S_{\rm N}$ and the projected distance from the cluster centre.
Models 1, 2, 3, 4, 5, and 6  are represented by
their numbers. 
The distance is defined as ($r_{\rm apo}$ + $r_{\rm peri}$)/2, where
$r_{\rm apo}$ and  $r_{\rm peri}$ are the apocenter and pericenter of  NGC 1404's orbit
in each model, respectively.
For comparison, the observed values of $S_{\rm N}$  and the projected distance  of the NGC 1404
from NGC 1399 is plotted by a cross.
}
\label{Figure. 6}
\end{figure}

\begin{figure}
\psfig{file=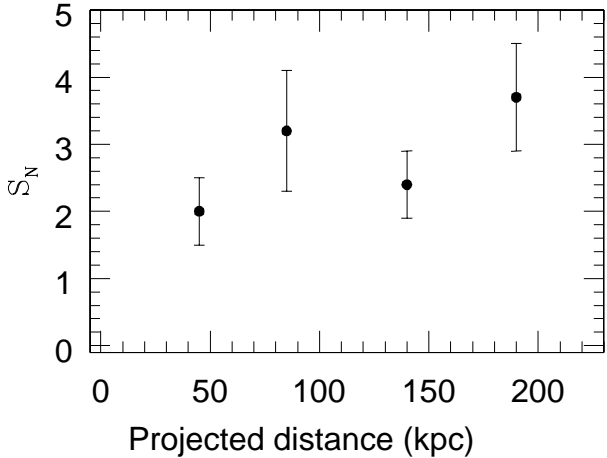,width=8.cm}
\caption{The observed variation of $S_{\rm N}$ with projected distance for ellipticals in
the Fornax cluster (Forbes et al. 1997). Here the projected distance is measured from the central
cD galaxy NGC 1399. The results are shown for NGC 1374, 1379, 1387, and 1404. 
}
\label{Figure. 7}
\end{figure}

\begin{figure}
\psfig{file=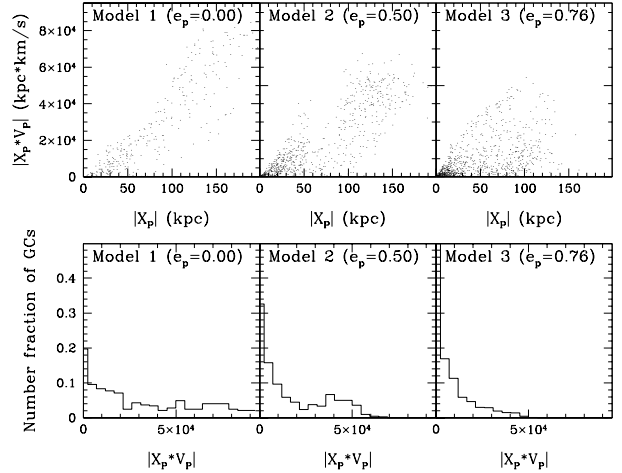,width=8.cm}
\caption{ 
$Upper$ $three$: Distributions of ICGCs 
on the $|X_{\rm p}|$-$|X_{\rm p}| \times |V_{\rm p}|$ planes,
where $|X_{\rm p}|$ and $|V_{\rm p}|$ represents the projected
distance  from the centre of Fornax cluster and velocity, respectively,
for the three representative models with different orbital
eccentricity ($e_{\rm p}$) of NGC 1404. 
The properties $|X_{\rm p}| \times |V_{\rm p}|$ can be considered
to correspond to the  ``projected orbital angular momentum'' of ICGCs. 
$Lower$ $three$: Number histogram of $|X_{\rm p}| \times |V_{\rm p}|$ of ICGCs
in the three models. For clarity, the number fraction of ICGCs in each bin
is shown. 
}
\label{Figure. 8}
\end{figure}

\begin{figure}
\psfig{file=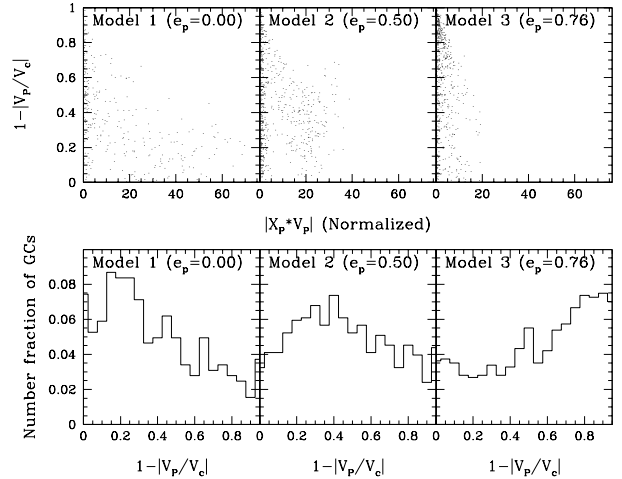,width=8.cm}
\caption{ 
$Upper$ $three$: Distributions of ICGCs 
on the $|X_{\rm p}| \times |V_{\rm p}|$-($1-|V_{\rm p}/V_{\rm c}|$) planes,
where $|X_{\rm p}|$,  $|V_{\rm p}|$, and $|V_{\rm c}|$ represent the projected
distance  from the centre of Fornax cluster, and the projected  velocity,
and the circular velocity estimated from the adopted NFW mass model
for Fornax cluster at $|X_{\rm p}|$, respectively,
for the three representative models with different orbital
eccentricity ($e_{\rm p}$) of NGC 1404. 
The properties $1-|V_{\rm p}/V_{\rm c}|$  can be considered
to correspond to the  ``projected orbital eccentricity'' of ICGCs. 
$Lower$ $three$: Number histogram of $1-|V_{\rm p}/V_{\rm c}|$ of ICGCs
in the three models. For clarity, the number fraction of ICGCs in each bin
is shown. 
}
\label{Figure. 9}
\end{figure}

\section{Discussion}

\subsection{Observable evidence for the tidal stripping scenario for low $S_{\rm N}$ 
cluster ellipticals} 

So far, we have  focused on just one elliptical galaxy (NGC 1404)
in Fornax cluster and the origin of its relatively low $S_{\rm
N}$. Based on these results, we propose that we can assess the
importance of tidal stripping by strong cluster gravitational
fields in the formation of low $S_{\rm N}$ elliptical galaxies
{\it in general} by checking the following three observable
physical properties of GC systems of cluster ellipticals. The
foremost is {\it the ratio of $S_{\rm N}$ within $1-2R_{\rm e}$ 
to that within 5--10 $R_{\rm e}$ }, where $R_{\rm
e}$ is the effective radius of a cluster elliptical galaxy. From
now on, we refer to this $S_{\rm N}$ ratio as $r_{\rm sn}$ just
for convenience. As our simulations have demonstrated, the outer
GCs are much more efficiently stripped than the inner GCs in an
elliptical galaxy. As a natural result of this, for example, the
ratio of $S_{\rm N}$ within  $R_{\rm e}$ to that within
10 $R_{\rm e}$ in NGC 1404 is changed from 0.18 into 0.49 during
2 Gyr of dynamical evolution (for an orbital eccentricity of
0.76). We can therefore  distinguish the tidal
stripping scenario of low $S_{\rm N}$ E formation from that in
which NGC 1404 has initially a small $S_{\rm N}$ of $\sim$ 2,
because only the former scenario predicts a larger  $r_{\rm
sn}$ ($>$ 0.2).  NGC 1404 has a local $S_{\rm N}$ of 0.5 at $R$ =
$1R_{\rm e}$ and 2.0 at $R$ = $9R_{\rm e}$ and thus a larger  $r_{\rm
sn}$ of 0.25, which implies that tidal stripping can be
responsible for the observed low $S_{\rm N}$ of 2 in NGC 1404.

The second observable characteristic of the tidal stripping
scenario is the formation of an elongated or flattened
distribution (or ``tidal stream'') of ICGCs along the orbit of
their previous host galaxy. We do, however,  point out that it is
formidable task to determine which cluster member galaxy
previously hosted each ``ICGC stream''.  Kinematic and
metallicity information of ICGCs may help up to identify each
individual stream, only if the ICGCs have not been affected by
scattering and dynamical relaxation via interaction between these
ICGCs and other cluster member galaxies since they were stripped
from their host galaxies. The central region of a cluster is
a place where many galaxies mutually interact with one another, and
much larger number of metal-poor GCs could be stripped from
numerous dwarf galaxies surrounding the central cluster cD, which
hampers the identification of the possible ICGC streams formed by
GC stripping from cluster ellipticals. Thus we conclude that it
is very hard to 
demonstrate that the
process of GC stripping is taking place from the projected distributions
alone.

The third is the statistical correlation between the distance of
an elliptical galaxy from the centre of a cluster and the $S_{\rm N}$
of the galaxy.  The present numerical results predict that if the
orbital eccentricities of galaxies does not depend on the
location of these with respect to the centre of a cluster,
$S_{\rm N}$ of a galaxy can correlate with the distance from the
cluster centre such that $S_{\rm N}$ is smaller for a galaxy
close to the cluster core.  Some evidence of this trend can be 
seen in the Fornax cluster. 
We expect the proposed 
correlation to be more obvious in the more dynamically
relaxed/older galaxy  clusters, because elliptical galaxies
in these clusters have already experienced pericenter passages
several times and lost their GCs by tidal stripping.  In our
future work we will generalise our simulations to more galaxies
within a cluster, and to clusters of varying properties.

\subsection{Kinematics of intracluster globular clusters as a new clue 
to galaxy evolution in clusters.}

Various different physical mechanisms are considered to play a 
role in galaxy evolution in clusters: e.g.,
Ram pressure stripping (Gunn \& Gott 1972),
tidal encounters (Iche 1985; Moore et al. 1996), 
tidal compression by the gravitational field of a cluster (Byrd \& Valtonen 1990),
minor or unequal-mass mergers (Bekki 1998).
One of the most important parameters which determine the effectiveness
of the each physical mechanism is suggested to be the pericenter
distance of the orbit of a galaxy. For example,
ram pressure stripping is efficient only when a gas-rich cluster galaxy
can pass through the cluster core where there is plenty of hot gas (e.g., Abadi et al. 1999).
Equally,  morphological transformation via a cluster tidal field can be significant 
only when the pericenter of a cluster galaxy is small enough to approach
the core radius (it should be noted here that
the latest high-resolution simulations on dynamical evolution of galaxies in
hierarchically forming clusters by Gnedin (2003)
demonstrate that the  tidal effects can be important throughout  the cluster). 
Therefore it is important  to  give some observational constraints
on the orbital properties  (e.g.,  mean eccentricity of  galaxy orbits and its dispersion)
of galaxies in clusters.

Based on the present numerical results, we propose that the {\it kinematical
properties of intracluster globular clusters (ICGCs) can provide 
some constraints on orbital populations of cluster galaxies and thus new clues
to the origin of galaxy evolution in clusters of galaxies}.  
Figs. 8 and 9 illustrate which kinematical properties can be useful
for this purpose.
In these two figures, we first choose GCs that are outside
10 $R_{\rm e}$ of NGC 1404 in each model and then investigated kinematical properties
of these stripped GCs that are regarded as as ICGCs drifting Fornax cluster. 
The upper  panel of Fig. 8 shows 
that the distribution of ICGCs on $|X_{\rm p}|$-$|X_{\rm p}| \times |V_{\rm p}|$ 
plane, where $|X_{\rm p}|$ and $|V_{\rm p}|$ are projected distance and velocity of a GC,
respectively, can provide some information on the orbital eccentricity ($e_{\rm p}$) 
of the previous host galaxy of ICGCs. 
ICGCs can be  distributed in the upper right region with larger $|X_{\rm p}| \times |V_{\rm p}|$ and
larger $|X_{\rm p}|$ in the low $e_{\rm p}$ model. 
The lower panel of Fig. 8 can provide
a clearer difference,
and demonstrates
that  the high  $e_{\rm p}$ model has the narrower  distribution in the histogram
of $|X_{\rm p}| \times |V_{\rm p}|$ and a larger number of
GCs in the smaller  $|X_{\rm p}| \times |V_{\rm p}|$ bins.
From these results, we  expect that if a cluster consists mostly of galaxies with 
higher orbital eccentricities, it shows narrower $|X_{\rm p}| \times |V_{\rm p}|$
histogram of ICGCs and their  ICGCs are located preferentially in the lower region
of the  $|X_{\rm p}|$-$|X_{\rm p}| \times |V_{\rm p}|$ plane of ICGCs. 
We can investigate the above just based on the projected distance and radial velocity
of each ICGC in a cluster.

Fig. 9 suggests that if we know the mass profile of a cluster and
thus its rotation curve through X-ray hot gas observations and
kinematical properties of cluster member galaxies, we can provide
even clearer evidence on orbital properties of cluster galaxies
by combining the ICGC kinematics and the cluster mass
profile. This figure demonstrates that if the orbits of cluster
galaxies are more eccentric as a whole, ICGCs (stripped from
these) populate mostly in the upper left region on the $|X_{\rm
p}| \times |V_{\rm p}|$-($1-|V_{\rm p}/V_{\rm c}|$) plane, where
$V_{\rm c}$ is the circular velocity of a ICGC at $|X_{\rm p}|$
and estimated from the adopted mass profile of a cluster.  This
is essentially because these ICGCs also have  highly eccentric
orbits and thus smaller projected velocity $|V_{\rm p}|$ for the
$V_{\rm c}$ at their positions.  Furthermore we can more clearly
see the difference between the models with different orbital
eccentricities in the histogram of $1-|V_{\rm p}/V_{\rm c}|$.
This  difference suggests that if the orbits of cluster
galaxies are more eccentric as a whole, such a cluster may show the
peak at high values of $1-|V_{\rm p}/V_{\rm c}|$. 
It should be noted here that the observed difference in
the above 
properties of ICGCs may not be clearly distinguished as proposed because of
 the mixture of different orbits of
galaxies in a cluster. However future 
observations on ICGCs in different nearby clusters of galaxies
can still  provide valuable constraints on galaxy evolution in
clusters.

\section{Conclusions}

We have numerically investigated the roles of the cluster tidal field
in dynamical evolution of the GC system of NGC 1404.  
We summarise our principle results as follows.

(1) Final $S_{\rm N}$ depends both on $e_{\rm p}$ (orbital
eccentricity of NGC 1404) and on $a_{\rm gc}$ (the ratio of the
scale length of the GC system to the effective radius of stellar
component of NGC 1404) in such a way that $S_{\rm N}$ is smaller
in the models with larger $e_{\rm p}$ and larger $a_{\rm gc}$.
$S_{\rm N}$ can be significantly  reduced by tidal stripping of GCs
to become as low as the observed value ($\sim$ 2), only if the
orbit of NGC 1404 is highly eccentric (with orbital eccentricity
of $>$ 0.5) and if the initial scale length of the GCs distribution is
about twice as large as the effective radius of NGC 1404.

(2) The radial number density profile of the GC system becomes
steeper after the stripping of GCs, because the outer GCs are
more easily stripped during tidal interaction. This result
implies that since the observed slope of the power-law density
profile of the GC system in NGC 1404 is shallower ($\sim$ $-1.3$)
compared with the typical value for cluster ellipticals ($-1.9$),
the initial slope (before tidal interaction) was rather shallow
(i.e. larger than $-1.3$), if the observed low $S_{\rm N}$ is due to
tidal stripping of GCs.

(3) One of the observable characteristics of a cluster elliptical
with a low $S_{\rm N}$ ($<$ 2) resulting from tidal
stripping is the larger ratio of $S_{\rm N}$
within $1-2R_{\rm e}$  to that within 5--10
$R_{\rm e}$, where $R_{\rm e}$ is the effective radius of a
cluster elliptical galaxy.  We also suggest that
if tidal stripping of GCs via a cluster tidal field is a main cause
for the evolution of $S_{\rm N}$ of cluster ellipticals, there
should be a positive correlation between the distance of a cluster
elliptical from the centre of a cluster and $S_{\rm N}$ of the
galaxy (as may be present in the Fornax cluster).

(4) Stripped GCs are found to become intracluster GCs (ICGCs)
orbiting the centre of Fornax cluster (i.e., NGC 1399) and their 
physical properties (e.g., number, radial distribution, and
kinematics with respect to the cluster centre) depend on the
orbit of NGC 1404 and the initial distribution of the GCs in NGC
1404.  For example, the ICGCs within the cluster core have
a projected number density profile with the power-law slope of
$\sim$ $-0.9$ and rather large velocity dispersion ($\sim$ 340 km
s$^{-1}$) for the highly eccentric orbit model of NGC 1404.

(5) Our numerical results suggest that not only structural
properties but  the kinematical properties  of ICGCs formed from
tidal stripping can depend strongly on the
orbits of their previous host galaxies. This implies that the 
detailed investigation of ICGC kinematics 
by multi-object spectrographs on 8-10m-class
telescopes can shed new insight into galaxy dynamics in the 
cluster as a whole. 

\section{Acknowledgment}
We are  grateful to the referee Oleg Gnedin for valuable comments,
which contribute to improve the present paper.
KB and WJC  acknowledge the financial support of the Australian Research Council
throughout the course of this work.
MB would like to thank the Swinburne Research and Development Grants Scheme.


\begin{figure}
\psfig{file=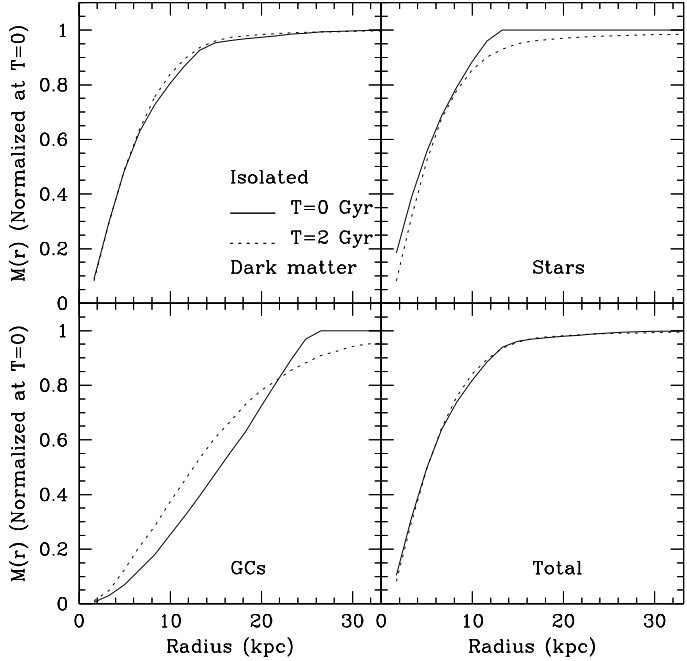,width=9.0cm}
\caption{ 
Cumulative mass distributions ($M(r)$) in the isolated model for
dark matter (upper left), stars (upper right), GCs (lower left),
and total (lower right) at $T$ = 0 Gyr (solid) and 2 Gyr (dotted).
$M(r)$ represents the total mass within $r$ (from the center of the model)
divided by the total mass of the model at $T$ = 0 Gyr for each component
(e.g, dark matter and GCs).
This normalized $M(r)$ in the isolated model enables us to estimate 
numerical relaxation effects on the radial distribution of each component.
}
\label{Figure. 10}
\end{figure}

\begin{figure}
\psfig{file=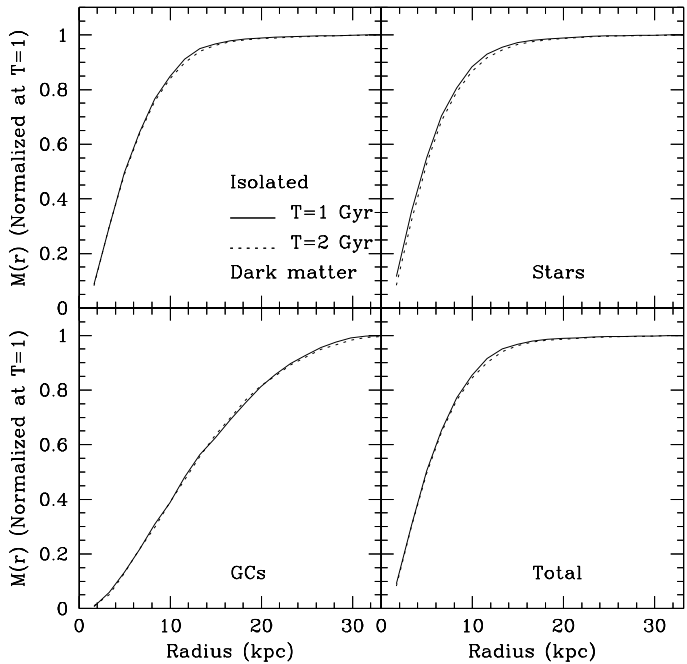,width=9.0cm}
\caption{ 
The same as Figure 10 but normalized at $T$ = 1 Gyr.
}
\label{Figure. 11}
\end{figure}

\begin{figure}
\psfig{file=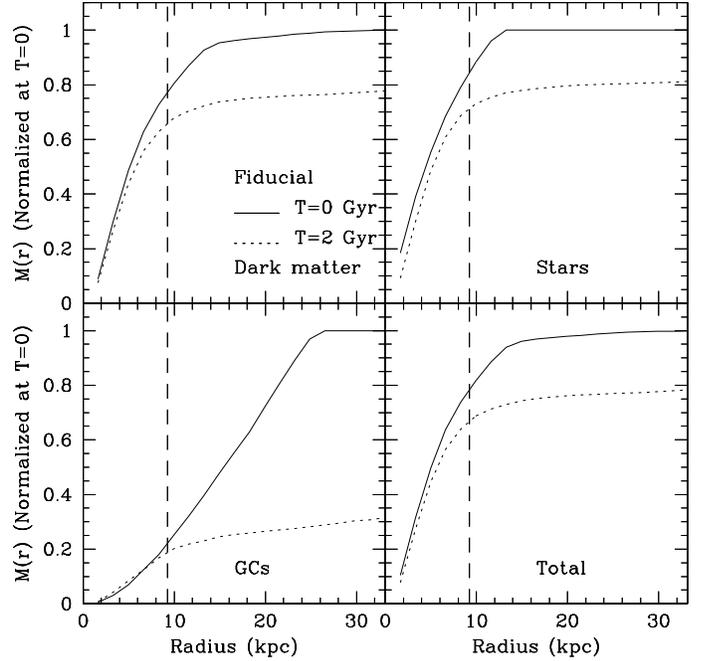,width=9.0cm}
\caption{ 
The same as Figure 10 but for the fiducial model.
For comparison, the tidal radius of this model is shown by a dashed line
in each panel. 
}
\label{Figure. 12}
\end{figure}

\appendix

\section[A]{Dependences on GC mass}

Although the main purpose of this paper is not to discuss
the dependences of tidal stripping processes 
on GC mass $M_{\rm gc}$, it is important
to confirm  whether the 
present numerical results  depend on  $M_{\rm gc}$.
We thus investigate the model with $M_{\rm gc}$ = $10^5$ $M_{\odot}$
(and 5 $\times$ $10^5$ $M_{\odot}$).
In these models with smaller $M_{\rm gc}$, the globular cluster system (GCS)  is more weakly
self-gravitating and the dynamical friction of each GC in the inner
part of NGC 1404 is less efficient (though the present simulations 
can not precisely treat 
the dynamical friction between GCs and the stellar and the  dark matter components 
of NGC 1404 because of the resolution of the simulations). 
Therefore, large  numbers of GCs can be stripped from NGC 1404 
during the tidal interaction between NGC 1404 and the Fornax cluster
in the model with smaller $M_{\rm gc}$.

The derived differences in final structural properties of the NGC 1404's GCS
between the two models with $M_{\rm gc}$ = $10^6$ $M_{\odot}$ (the fiducial model)
and $M_{\rm gc}$ = $10^5$ $M_{\odot}$ are summarized as follows.
Firstly, $R_{0.1}$, which is the radius
(from the centre of NGC 1404) within which 10 \%  of initial GCs  in number (i.e., 135) can be
located, is larger by 21 (23) \% in  the model with $M_{\rm gc}$ = $10^5$ $M_{\odot}$
than in the fiducial model at $T$ = 0.47 (0.94) Gyr.
This suggests that the final GCS  
has the lower number density in the central region of NGC 1404 for the model with
smaller $M_{\rm gc}$.
Secondly, the total number of stripped GCs (estimated for $R/R_{\rm e}$ $<$ 10)
is larger by 19 \% (22 \%) in  the model with $M_{\rm gc}$ = $10^5$ $M_{\odot}$ 
than in the fiducial model at $T$ = 0.47 (0.97) Gyr. 
This means that a larger number of GCs from NGC 1404 can form the ICGC population
in the Fornax cluster for the model with smaller $M_{\rm gc}$.
From these two results,  we can conclude that time evolution of $S_{\rm N}$,
final structural properties of the GCS in NGC 1404, and the total number of ICGC
originated from NGC 1404 in the Fornax cluster does not depend so strongly
on $M_{\rm gc}$. 
 
\section[B]{Long-term evolution of an isolated model and numerical relaxation effects}

It is possible that our numerical simulations overestimate the total number of stripped
GCs in  a cluster owing to the undesirable numerical relaxation effects from the
limited particle number used in the simulations.
Our numerical results could  be influenced by (1) the somewhat abrupt truncation of
mass distribution for each component and (2)  the difference in masses between
different components.
In order to estimate these purely numerical effects (and to discriminate
these effects with those from a cluster tidal force),
we investigate the long-term dynamical evolution
($\sim$ 2 Gyr corresponding to 4000 time steps)
of an isolated galaxy model with no external tidal force.
Figs. 10 and 11 show the mass distribution for each collisionless component
at $T$ = 0 and 2 Gyr in the isolated model.
Although the entire mass distribution of the isolated model does not
change so significantly during 2 Gyr evolution, the mass distributions of 
stars and GCs change slightly.
The stellar component shows  a small degree of expansion (i.e., lower density
in the inner and the outer regions), which may be caused by 
mass segregation  between less massive  stellar particles
and more massive dark matter ones (in particular, for the first 1 Gyr evolution).
Also the initial abrupt truncation of the stellar mass distribution may  be
responsible for the appreciable expansion.

The GC component at $T$ = 2 Gyr
shows the higher density in its inner region and the more diffuse
outer one compared with the initial distributions ($T$ = 0 Gyr).
This implies that owing to (1)  the  low GC number density in the
inner part and (2) the adopted large softening length of the GC particle,
the mass segregation   between less massive
GC particles and the more massive dark matter ones
does not proceed significantly  within a time scale of 2 Gyr
(compared with the case of the stellar component).
This also implies that the initial abrupt truncation of the GC (and stellar) distribution(s)
can cause a more serious numerical effect on dynamical evolution of the GC component
compared with the stellar component.
The GC component is  more widely distributed than the stellar one
and the initial velocity dispersion of the GC component is determined
by the local potential energy at the positions of GC particles.
Accordingly, the velocity dispersions  
of the outer GCs  ($R$ $>$ $5 r_{\rm e}$ $\sim$ 12.5 kpc),
which constitute the majority of GCs ($\sim$ 70 \%) in the isolated model,
must be determined by much  smaller  number of particles at $R$ $>$ $5 r_{\rm e}$
(where there are initially no stellar particles because of the adopted abrupt
truncation of the stellar distribution).
This less precise estimation of initial velocity dispersions
of the outer GCs (thus initially less stable dynamical state  of the GCs)
can be responsible for the dynamical relaxation 
that leads to  the formation of the ``core-halo'' structure 
(i.e., higher density inner region and the lower density outer one) seen in 
the mass distribution of the GC component at $T$ = 1 and 2 Gyrs.

The derived ``apparent'' decrease of GCs in the isolated model is at most
$5-10\%$ in the outer region of the model. Therefore, it is less likely that
we overestimate significantly the number of GCs stripped by the tidal force
of a cluster because of  the purely numerical effects described above in the
isolated model.
Fig. 12 describes the mass distribution of each component in the fiducial
model for the long-term (2 Gyr) dynamical evolution. 
It is clear from this figure that about 70 \% of GCs are stripped by
the clusters global tidal field during 2 Gyr. The derived number of
70 \% is much higher than that of $5-10\%$ which  is
derived for the isolated model and could be due to purely numerical
relaxation effects of the present simulations. 
Therefore, it can be concluded that numerical relaxation effects do not
significantly alter the present results described in the main text.  
Fig. 12 also clearly demonstrates  that the total number of stripped GCs
becomes suddenly larger beyond the tidal radius that is analytically estimated
for the NGC 1404 model  (see the main text).


\begin{thebibliography}{99}


\bibitem{ab99}
Abadi, M. G., Moore, B.,  Bower, R. G. 1999, MNRAS, 308, 947

\bibitem{}
Aguilar, L., Hut, P., Ostriker, J. P. 1988, ApJ, 335, 720


\bibitem{}
Arnaboldi, M., Freeman, K.C., Cappaccioli, M., Ford, H., 1994, ESO Messenger, 76, 40

\bibitem{az92}
Ashman, K. M.,  Zepf, S. E., 1992, ApJ, 384, 50



\bibitem{}
Barnes, J.E., 2000, astro-ph/0010145


\bibitem{}
Bassino, L. P., Cellone, S. A., Forte, J. C., Dirsch, B., 2002, in preprint
(astro-ph/0212532)




\bibitem{bm02} 
Beasley, M. A., Baugh, C. M., Forbes D. A., Sharples, R. M.,
Frenk, C. S. 2002, MNRAS, 333, 383


\bibitem{be98}
Bekki, K. 1998, ApJ, 502, L133

\bibitem{}
Bekki, K., Forbes, D. A., Beasley, M. A.,  Couch, W. J. 2002, MNRAS, 335, 1176

\bibitem{}
Bicknell, G. V., Bruce, T. E. G., Carter, D.,   Killeen, N. E. B.
1989, ApJ, 336, 639

\bibitem{}
Binney, J.,  Tremaine, S. 1987 in Galactic Dynamics, Princeton; Princeton
Univ. Press.

\bibitem{}
Blakeslee, J. P., Tonry, J. L., Metzger, M. R., 1997, AJ, 114,
482


\bibitem{by90}
Byrd, G.,  Valtonen, M. 1990, ApJ, 350, 89

\bibitem[]{}  C\^ote,~P., Marzke,~R.~O., West,~M.~J., 1998, ApJ, 501, 554 

\bibitem{}
Dirsch, B., Richtler, T., Geisler, D., Forte, J. C., Bassino, L. P.,
Gieren, W. P. 2003, AJ, 125, 1908

\bibitem{}
Fabian, A., Nulsen, P., Canizares, C., 1984, Nature, 310, 733

\bibitem{fo97} Forbes, D. A., Brodie, J. P., Grillmair, C. J., 1997, AJ, 113, 1652  

\bibitem{fo01} Forbes, D. A., 2001, preprint (astro-ph/0106040) 

\bibitem{}
Forte, J. C.; Martinez, R. E.,   Muzzio, J. C., 1982, AJ, 87, 1465

\bibitem{}
Forte, J. C., Geisler, D. K. E.,  Lee, M. G.,
Ostrov, P. 2002, in Extragalactic star clusters 207th IAU symposium,
edited by D. Geisler, E. K. Grebel, and D. Minniti, p251  

\bibitem{}
Geisler, D.,  Lee, M. G.,   Kim, E. 1996, AJ, 111, 1529

\bibitem{}
Gnedin, O. Y. 2003, ApJ, 582, 141


\bibitem{}
Grillmair, C.J., Freeman, K.C., Bicknell, G.V., Carter, D., Couch, W.J., 
Sommer-Larsen, J.,  Taylor, K., 1994, ApJ, 422, 9

\bibitem{gg72}
Gunn, J. E.,  Gott, J. R., III 1972, ApJ, 176, 1




\bibitem{ha91} Harris, W. E., 1991, ARAA, 29, 543 

\bibitem{}
Harris, W., 2001, in Star Clusters, Saas-Fee Advanced Course 28, 
edited by L Labhardt and B. Binggel, p223 

\bibitem{}
Harris, W., Harris, G.,   McLaughlin, D., 1998, AJ, 115, 1801

\bibitem{}
Hilker, M., Infante, L.,   Richtler, T. 1999, A\&AS, 138, 55

\bibitem{ic85}
Icke, V. 1985, A\&A, 144, 115

\bibitem{jo97}
Jones, C., Stern, C., Forman, W., Breen, J., David, L., Tucker, W., 
Franx, M. 1997, ApJ, 482, 143

\bibitem{ki99} Kissler-Patig, M.,; Grillmair, C. J., Meylan, G.,
Brodie, J. P., Minniti, D.,  Goudfrooij, P., 1999, AJ, 117, 1206


\bibitem{mac99} McLaughlin, D. E., 1999, AJ, 117, 2398

\bibitem{}
Minniti, D., Kissler-Patig, M., Goudfrooij, P., Meylan, G., 1998, AJ, 115, 121

\bibitem{mo96}
Moore, B., Katz, N., Lake, G., Dressler, A.,  Oemler, A., Jr., 
1996, Nature, 379, 613


\bibitem{}
Muzzio, J. C.; Martinez, R. E.; Rabolli, M., 1984, ApJ, 285, 7

\bibitem{}
Muzzio, J. C., 1986, ApJ, 301, 23

\bibitem{}
Muzzio, J. C., 1986, ApJ, 306, 44

\bibitem{}
Muzzio, J. C.; Dessaunet, V. H., Vergne, M. M., 1987,
ApJ, 313, 112

\bibitem{}
Muzzio, J., 1987, PASP, 99, 245

\bibitem{}
Napolitano, N.R., Arnaboldi, M., Capaccioli, M., 2002, A\&A, 383, 791

\bibitem{na96}
Navarro, J. F., Frenk, C. S.,  White, S. D. M
1996, ApJ, 462, 563

\bibitem{}
Paolillo, M., Fabbiano, G., Peres, G., Kim, D.W., 2002, ApJ, 565, 883


\bibitem{sug90}
Sugimoto,~D., Chikada,~Y., Makino,~J., Ito,~T., Ebisuzaki,~T., 
Umemura, M. 1990, Nature, 345, 33

\bibitem{}
West, M., et al. 1985, ApJ, 453, L77

\bibitem{}
White, R. E. III, 1987, MNRAS, 227, 185



\bibitem{ze96} Zepf, S. E., Ashman, K. M., English, J.,
Freeman, K. C.,    Sharples, R. M., 1999, AJ, 118, 752 


\end{thebibliography}
\end{document}